\documentclass[aps,prx,superscriptaddress,twocolumn,longbibliography]{revtex4-1}

\usepackage{amsfonts}
\usepackage{subfigure}
\usepackage{amsmath}
\usepackage{txfonts}
\usepackage{amssymb}
\usepackage{amsbsy} 
\usepackage{epsfig}
\usepackage{graphicx}
\usepackage{float}
\usepackage{epstopdf}

\usepackage{float} 

\def\be{\begin{equation}} \def\ee{\end{equation}}
\def\bea{\begin{eqnarray}} \def\eea{\end{eqnarray}}
\def\bw{\begin{widetext}} \def\ew{\end{widetext}}

\def\bk{{\bf k}}

\def\be{{\bf e}}

\def\bS{{\bf S}}
\def\bR{{\bf R}}

\def\hs{\hat s}

\def\to{\tilde \epsilon}
\def\tr{\tilde \rho}

\def\tz{\tilde z}
\def\tg{\tilde g}
\def\tR{\tilde R}
\def\tG{\tilde{\mathcal G}}

\def\rw{\rightarrow}

\begin{document}

\title{RKKY interaction of magnetic impurities in multi-Weyl semimetals}

\author{Yong Sun}
\altaffiliation{ suninsky@mail.ustc.edu.cn} \affiliation{ Department of Modern Physics, University of Science and Technology of China, Hefei, China, 230026}

\author{Anmin Wang}
\affiliation{ Department of Modern Physics, University of Science and Technology of China, Hefei, China, 230026}

\begin{abstract}
We have systematically investigated the Ruderman-Kittel-Kasuya-Yosida (RKKY) interaction between two magnetic impurities in Weyl semimetals with
arbitrary monopole charge $Q$. We find that the RKKY interaction becomes intrinsically anisotropic for
$Q\geq2$, and its dependence on Fermi energy and impurity separation is directly controlled by the monopole charge.
With the increase of $Q$, the RKKY interaction becomes more long-ranged and more anisotropic, which makes
interesting magnetic orders easier to form and thus may have important applications in spintronics. 
\end{abstract}

\date{\today}
\maketitle

\section{Introduction}\label{intro}

Electronic band structure plays a fundamental role in condensed matter physics
because many properties of materials are directly determined by it~\cite{kittel2005introduction}.
Recently, Weyl semimetals (WSMs) have attracted major research interest because
their energy bands host isolated crossing points protected by topology~\cite{Murakami2007weyl,wan2011weyl,Burkov2011weyl,
Weng2015weyl,huang2015weyl,Xu2015weyl,Lv2015weyl,lu2015weyl}.
A remarkable property of the crossing points is that they
carry integer Berry flux and thus can be taken as monopoles in momentum space~\cite{Xiao2010berry}.
As a direct consequence of the integer Berry flux (or the monopole charge), equal numbers of 
chiral Landau levels will
show up in the presence of a magnetic field. If an electric field is further
applied in parallel with the magnetic field, novel phenomena
related to chiral anomaly will take place~\cite{nielsen1983adler,Son2013anomaly,
Huang2015anomaly,Parameswaran2014anomaly,Zhang2016anomaly}.

Although WSMs have been extensively studied, most works are restricted to the WSMs
belonging to the class with monopole charge $Q=1$~\cite{yan2017review,Armitage2017review,Burkov2017review,Hasan2017review}, 
and the classes with  $Q\geq2$ remain much less explored due to lack of experimentally confirmed materials. 
In contrast with the $Q=1$ class, the classes with $Q\geq2$
will inevitably show anisotropic energy dispersion away from the crossing points~\cite{xu2011,Fang2012mwsm,huang2016}.
Consequently, their density of states will also show distinctive power laws, which immediately
indicates that the monopole charge will affect a series of physical properties~\cite{Lai2015dwsm,Jian2015dwsm,
Ahn2016mwsm,Ahn2017mwsm,Park2017mwsm,Zhang2016mwsm,Wang2017mwsm,Sun2017dwsm,Huang2017mwsm}. 
Motivated by these observations, in this work we give a systematic study on the Ruderman-Kittel-Kasuya-Yosida
(RKKY) interaction~\cite{Ruderman1959rkky,kasuya1956rkky,Yosida1957rkky}, which is an indirect interaction between magnetic impurities
induced by itinerant carriers,  for WSMs with arbitrary monopole charge.

The RKKY interaction in WSMs with $Q=1$ have already been investigated~\cite{chang2015,hosseini2015}. Based on the ideal
isotropic model, the authors found that the interaction is isotropic, and for the intrinsic case it
 decays with a power law $H_{\text{RKKY}}(R)\propto R^{-5}$ with $R$ the distance between two impurities; 
while for finite doping, 
the decaying power law becomes $H_{\text{RKKY}}(R)\propto R^{-3}$, indicating that the interaction 
is in fact quite short-ranged~\cite{chang2015,hosseini2015}. In this work, we generalize these studies 
to arbitrary large $Q$, and find that for $Q\geq2$,
the interaction becomes anisotropic, with $H_{\text{RKKY}}(z)\propto z^{-4/Q-1}$ and
$H_{\text{RKKY}}(\rho)\propto \rho^{-Q-4}$ for the intrinsic case, and 
$H_{\text{RKKY}}(z)\propto z^{-2/Q-1}$ and $H_{\text{RKKY}}(\rho)\propto \rho^{-3}$
for finite doping. Thus, for large $Q$, the interaction becomes quasi-one-dimensional
and long-ranged, which may trigger interesting magnetic orders. Besides, 
the power law can be utilized as a way to determine the monopole charge of WSMs.

The paper is organized as follows. In Sec.~\ref{ssetup}, we outline the setup of our model and deduce general forms of RKKY range functions. In Sec.~\ref{results}, we present long-range asymptotic results for two most representative alignments of impurities, namely along the line connecting the multi-Weyl points and in the perpendicular plane. Both the dependence on impurity separation and on Fermi energy are discussed. We end with brief conclusion in Sec.~\ref{concl}.

\section{The Setup}\label{ssetup}

As Weyl points emerge from the touching of two adjacent non-degenerate bands, the low-energy effective Hamiltonian near Weyl points with monopole charge $Q$ can be generally written down as follows,
\begin{equation}
H_{0}(\bk)= \chi \lambda\left(k_- ^Q \sigma_+ + k_+^Q \sigma_- \right)
+ \chi v (k_z-\chi k_0) \sigma_z, \label{H0}
\end{equation}
in which $\sigma_\pm=\frac{1}{2} (\sigma_x \pm i \sigma_y)$ and $k_\pm = k_x \pm i k_y$, $\sigma_{x,y,z}$ denote Pauli matrices in real spin space; $\chi = \pm$ denote two kinds of chirality of multi-Weyl points; $\lambda$ is a parameter with mass dimension $(1-Q)$, and $v$ is the Fermi velocity in the $z$ direction. The multi-Weyl points are located at $\pm(0,0, k_0)$ in momentum space. We will refer to the line connecting the multi-Weyl points as the vertical direction, and the plane perpendicular to the line as the transverse plane.

Now we consider two magnetic impurities with localized spins ${\bf S}_{1}$ and ${\bf S}_{2}$ that are well embedded into the WSMs, such that the effect of surface states can be neglected. Besides, for simplicity of notation, the two localized spins are
placed at the origin of coordinates and position $\bf R$. The standard $s$-$d$ interaction,
which describes the coupling between localized spins and itinerant electrons, is given by
\begin{eqnarray}
H_I = (J\tau_0+\Lambda \tau_x)\mathbf{S}_i\cdot\boldsymbol{\sigma}\delta(\mathbf{r}-\mathbf{R}_i),
\end{eqnarray}
where $J$ and $\Lambda$ represent the intranode and internode coupling strength, respectively.
The $s$-$d$ interaction can be treated as a perturbation to the multi-Weyl Hamiltonian.
At zero temperature, the RKKY interaction between these two magnetic impurities can be
obtained by second order perturbation theory~\cite{Ruderman1959rkky,kasuya1956rkky,Yosida1957rkky,chang2015,hosseini2015}, which is
\begin{eqnarray}
&& H_{\text{RKKY}} = \sum\limits_{\alpha,\beta,\chi,\chi^{\prime}}\left[J^{2}\delta_{\chi\chi^{\prime}}+\Lambda^{2}
(1-\delta_{\chi\chi^{\prime}})\right]S_{1}^{\alpha}S_{2}^{\beta}\nonumber\\
&& \times\mathrm{Im} \left\{ -\frac{1}{\pi}\int_{-\infty}^{\epsilon_{F}}d\epsilon
\mathrm{Tr}\left[\sigma_{\alpha}G_{\chi} (\epsilon,\bf R)\sigma_{\beta} G_{\chi^{\prime}} (-\epsilon,\bf R)\right]\right\},\label{RKKY}
\end{eqnarray}
where $\epsilon_F$ is the Fermi energy, $G_{\chi}(\epsilon,\bf R)$ denotes the real-space Green's function matrix in the absence
of magnetic impurities.

\subsection{GREEN'S FUNCTION}

In the absence of magnetic impurities, the Green's function in momentum space takes
the form of $G_\chi^{-1}(\epsilon,\mathbf k)=\epsilon-H_{0}(\mathbf k)$. To get its form in
the energy-coordinate representation, we perform a Fourier transformation,
\begin{equation}
G_\chi(\epsilon,\mathbf R)=\int \frac{d^3 k}{(2\pi)^3}\frac{\epsilon+\tilde{H}_{0}(\mathbf k)}{\epsilon^2-E_k^2}e^{i\mathbf k \cdot \mathbf R} e^{i \chi k_0 z},
\end{equation}
where $E_k = \sqrt{{\lambda}^2 k_\rho^{2 Q}+v^2k_z^2}$ with $k_\rho=\sqrt{k_{x}^2+k_{y}^2}$, and
$\tilde{H}_{0}(\bk)=\chi \lambda\left(k_- ^Q \sigma_+ + k_+^Q \sigma_- \right)
+ \chi v k_z\sigma_z$. Note that the energy dispersion relation is linear in the vertical direction and non-linear in the transverse plane for $Q\geq2$.

The above integration can be more conveniently solved in spherical coordinates, i.e. $(k_x,k_y,k_z)\rightarrow(k,\phi,\psi)$, where $\phi$ denotes the angle between the momentum vector and the $x$-$y$ plane. Also, we write $\bR$ in cylindrical coordinate ${\bf{R}} = (\rho,\theta,z)$. After some straightforward calculations, we find that
\begin{eqnarray}
    G_{\chi}(\epsilon,\bR) =  C_g e^{i \chi k_0 z}  \left(g_0 \sigma_0 +\chi g_1 \sigma_s + \chi g_3 \sigma_z  \right),\label{gwrpl}
\end{eqnarray}
where $\sigma_s = {\bf{\sigma}} \cdot {\hat s}(\theta)$, with $\hs(\theta)=\cos(Q \theta) \hat x + \sin(Q \theta) \hat y$, and the constant multiplier defined as $C_{g}=-\frac{1 }{4 Q \pi^2 v} \left (\frac{v k_0}{\lambda}\right)^{2/Q}$.
The dimensionless coefficients $g_{0,1,3}(\epsilon,\rho,z)$ are complex functions of the following form,

\bw
\bea
g_0(\epsilon,\rho,z) &=&  \frac{\epsilon}{ (k_0 v)^{2/Q}} \int_0^{+\infty}dq \int_{-\pi/2}^{\pi/2}d\phi \frac{ q^{2/Q}   (\cos \phi)^{2/Q-1}  J_{0} \left(\left( q \cos\phi/\lambda \right)^{1/Q} \rho \right) \cos \left( q z \sin\phi/v \right)           }{q^2-\epsilon^2},\nonumber\\
g_1(\epsilon,\rho,z) &=&  \frac{i^Q}{ (k_0 v)^{2/Q}}    \int_0^{+\infty}dq \int_{-\pi/2}^{\pi/2}d\phi \frac{ q^{2/Q+1} (\cos \phi)^{2/Q  }  J_{Q} \left(\left( q \cos\phi/\lambda \right)^{1/Q} \rho \right) \cos \left( q z \sin\phi/v \right)           }{q^2-\epsilon^2},\nonumber\\
g_3(\epsilon,\rho,z) &=&  \frac{i}{ (k_0 v)^{2/Q}}      \int_0^{+\infty}dq \int_{-\pi/2}^{\pi/2}d\phi \frac{ q^{2/Q+1} (\cos \phi)^{2/Q-1}  J_{0} \left(\left( q \cos\phi/\lambda \right)^{1/Q} \rho \right) \sin \left( q z \sin\phi/v \right) \sin\phi  }{q^2-\epsilon^2}.
\eea
\ew

$G_{\chi}(\epsilon,-\mathbf R)$ can be readily obtained by substituting $\theta \rw \theta+\pi$ and $z\rw-z$ into Eq.~(\ref{gwrpl}). Noticing that $\hs(\theta+\pi)=(-1)^Q \hs(\theta)$, we readily have

\begin{eqnarray}
    G_{\chi}(\epsilon,-\bR) = C_g e^{-i \chi k_0 z} \left(g_0\sigma_0 + (-1)^Q \chi g_1\sigma_x - \chi g_3 \sigma_z  \right).\label{gwrmi}
\end{eqnarray}
Bringing these results back into Eq.(\ref{RKKY}), the composition of RKKY interaction can now be analyzed.

\subsection{RKKY INTERACTION}

We find that the RKKY interaction in multi-WSMs generally have four distinctive types of terms,
\begin{eqnarray}
    H_{\mathrm{RKKY}}\left(\epsilon_F,\bR \right) &=& F_1 \mathbf S_1 \cdot \mathbf S_2 + F_2 \left(\mathbf S_1 \times \mathbf S_2\right)\cdot\hs \nonumber \\
    && + F_3 (\bS_1\cdot\hs) (\bS_2\cdot\hs) + F_4 S_1^z S_2^z, \label{rkky4}
\end{eqnarray}
where the range functions $F_{1,2,3,4}$ are defined as follows:
\bea
    F_1 &=& \frac{1}{2} \bigg[ \mathcal G_{00} \left(J^2 + \Lambda^2 \cos (2 k_0 z) \right) \nonumber\\
    &&+ \left(\mathcal G_{33}+(-1)^{Q+1}\mathcal G_{11}\right)\left(J^2 - \Lambda^2 \cos (2 k_0 z)\right) \bigg], \nonumber\\
    F_2 &=& \frac{1+(-1)^Q}{2} \mathcal G_{01} \Lambda^2 \sin (2 k_0 z), \nonumber\\
    F_3 &=&   (-1)^Q\mathcal G_{11} \left( J^2-\Lambda^2 \cos (2k_0 z)\right ), \nonumber\\
    F_4 &=& - \mathcal G_{33} \left( J^2-\Lambda^2 \cos (2k_0 z)\right ).\label{rangefunc}
\eea
In the above equation, the first term is the rotation-invariant Heisenberg term, which 
favors either parallel or antiparallel alignment of the impurity spins
depending on the sign of $F_{1}$. The second term is the Dzyaloshinsky-Moriya (DM) 
term~\cite{dzyaloshinsky1958dm,Moriya1960DM}, which favors configurations in which the impurity
spins are orthogonal; this term should vanish for odd $Q$ because inversion symmetry remains 
intact in these cases. The third term is the so-called spin-frustrated term, which favors the 
(anti)parallel alignment of the impurity spins along $\hs$. Importantly, for both the DM and 
spin-frustrated terms, the favored direction $\hs$ generally does not coincide with $\hat \rho$ 
(alignment of the two magnetic impurities projected on the transverse plane) for $Q\geq 2$, which 
predicts novel spin structures for multi-WSMs. The last term is the Ising term. Note 
that when $Q=1$, i.e. for single Weyl nodes, $\hat s = \hat \rho$ and Eq.~(\ref{rkky4}) reduces 
to the form derived in~\cite{hosseini2015}.

Note that when we generalize our system to many magnetic impurities which are distributed randomly in real materials, the spin-frustrated term frustrates the spins of magnetic impurities, hence the terminology. Furthermore, the rapid oscillating terms $\Lambda^2 \cos (2 k_0 z)$ and $\Lambda^2 \sin (2 k_0 z)$ 
should average out (over impurity positions) for large momentum separation $2k_0$ and do not 
contribute to net magnetization. The intranode process thus contribute predominately to the 
RKKY interaction~\cite{chang2015,brey2007}. We will take this simplification in the following discussions.

Our main quest now reduces to calculating the $\mathcal G_{ij}$ coefficients, as defined for $i,j\in\{0,1,3\}$,
\begin{eqnarray}
    \mathcal G_{ij} (\epsilon_F,\bR) &\equiv& \frac{C}{k_0 v} \times \mathrm{Im} \int_{-\infty}^{\epsilon_{F}} d\epsilon g_i g_j\nonumber\\
    &=&C \mathrm{Im} \int_{-\infty}^{\tilde\epsilon_{F}} d\tilde\epsilon \tilde g_i \tilde g_j \equiv C \tilde{\mathcal G}_{ij},
\end{eqnarray}
with constant $C= -\frac{k_0}{2 Q^2 \pi^5 v} \left( \frac{v k_0}{\lambda} \right)^{4/Q}$. The momentum separation of multi-Weyl points provides a natural scale for our system. For convenience, we will switch to dimensionless variables $\tilde \epsilon \equiv \epsilon/(k_0 v)$, $\tilde z \equiv k_0 z$ and $\tilde \rho \equiv (k_0 v/\lambda)^{1/Q} \rho$. Naturally, two most interesting cases are $\tr=0$ and $\tz=0$, which correspond to impurities alignment along the $z$ axis and in the transverse plane, respectively. We discuss both scenarios in details below. The $\mathcal G_{ij}$'s are generally not analytically integrable. Fortunately, for large values of $\tz$ and $\tr$, we manage to devise analytical approximations that agree well with numerical results.

\section{RESULTS AND DISCUSSIONS}\label{results}

\subsection{IMPURITIES AlONG THE VERTICAL DIRECTION}\label{szaxis}

\begin{figure}
\includegraphics[width=7cm, height=11cm]{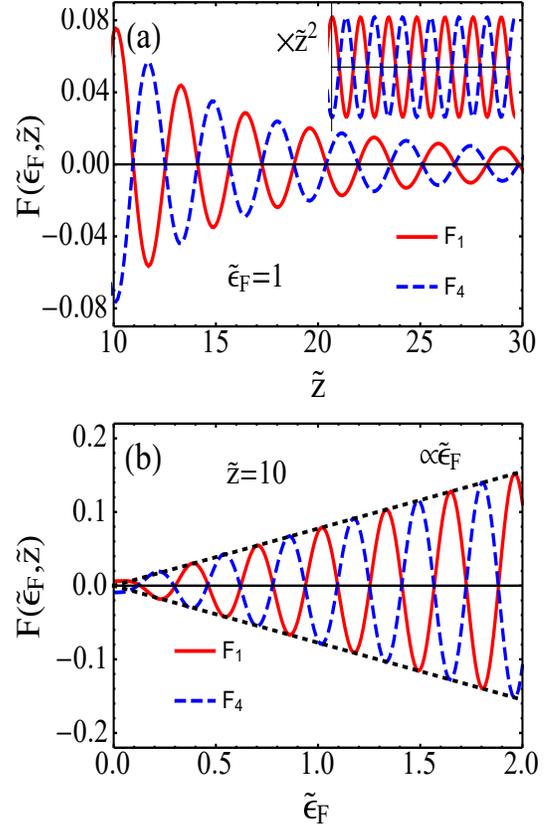}
\caption{The exact RKKY range functions  (in units of $J^2 C$) for impurities in the vertical direction. Here $Q=2$. The top and bottom panels show power law dependence of RKKY range functions on impurity separation and on Fermi energy, respectively. The Ising term cancels with the $z$-th component of the Heisenberg term, leading to $XY$-like spin model.} \label{zplot}
\end{figure}

For magnetic impurities aligned along the vertical direction, i.e., $\tr=0$, as $J_Q(0)=0$ for $Q \in \mathbb Z^+$ and $J_0(0)=1$,
the $\tg_i$ coefficients can be simplified to
\bea
\tg_0 &=&  \int_0^{+\infty}dq \int_{-\pi/2}^{\pi/2}d\phi \frac{\tilde \epsilon q^{2/Q}   (\cos \phi)^{2/Q-1}  \cos \left(  q \tilde z \sin\phi \right)           }{q^2-\tilde \epsilon^2},\nonumber\\
\tg_3 &=&  i      \int_0^{+\infty}dq \int_{-\pi/2}^{\pi/2}d\phi \frac{ q^{2/Q+1} (\cos \phi)^{2/Q-1}  \sin \left( q \tilde z \sin\phi \right) \sin\phi  }{q^2-\tilde \epsilon^2},\nonumber\\
\tg_1 &=&  0.\label{gz}
\eea
\
The asymptotic dependence of $\tg_i$ on $\tilde z$ and $\tilde \epsilon$ can be shown heuristically. For $\tg_0$, the integration of the numerator over $\phi$ has $\to (q /\tz)^{1/Q} \cos(q\tz-\pi/2Q)$ as the leading term, the subsequent Cauchy integration over $q$ gives $\tg_0 \propto (\to/\tz)^{1/Q}\exp(i (\to \tz -\pi/2Q) )$. Also, it is explicit from Eq.~(\ref{gz}) that $\tg_3 = -\frac{i}{\to} \frac{\partial}{\partial \tz}\tg_0$. At large distance $\tz >>1$, it is readily seen that $\tg_0\approx\tg_3$.

The coefficients $\tG_{ij}$ thus take the following form for finite Fermi energy and at long distance,
\begin{equation}
    \tG_{00}(\to_F,\tz)\approx\tG_{33}(\to_F,\tz)\approx  \frac{\alpha_1\to_F^{2/Q}}{\tz^{2/Q+1}}\cos(2\to_F\tz-\frac{\pi}{Q}),\label{G00z}
\end{equation}
where $\alpha_1=\frac{2^{2/Q}\pi^2}{8} \left(\Gamma\left( \frac{1}{Q} \right)\right)^2$ is a real constant that only depends on chiral charge $Q$.

It is understood that the integration over occupied states in the valence band ($\int_{-\infty}^0 d \to \tg_i^2$) generates unphysical divergence, which can be regulated using the soft cutoff procedure~\cite{saremi2007}. Specifically, for the intrinsic case ($\epsilon_F=0$), the corresponding form can be obtained by dimensional analysis,
\begin{equation}
    \tG_{00}(0,\tz)\sim\tG_{33}(0,\tz)\propto\frac{1}{\tz^{4/Q+1}}.\label{G00zi}
\end{equation}
The above terms could be safely dropped from Eq.~(\ref{G00z}) since we are only interested in the long range scenario. Note that the leading term argument does not apply to infinite integrals, hence the $\sim$ between $\tG_{00}(0,\tz)$ and $\tG_{33}(0,\tz)$ indicating different constant multipliers. The constant multipliers can in principle be worked out numerically for each chiral charge $Q$.

Finally, we have the long range asymptotic form of RKKY interaction for impurities along $z$ axis. For finite doping,
\bea
H_{\mathrm{RKKY}}^z & \propto & \frac{\to_F^{2/Q}\cos(2\to_F\tz-\pi/Q)}{\tz^{2/Q+1}} (\bS_1 \cdot \bS_2 - S_1^z S_2^z). \label{hz}
\eea

For reference, Fig.~\ref{zplot} plots exact numerical results of RKKY range functions in double-Weyl semimetal ($Q=2$) with vertical impurity alignment. The power law dependence on impurity separation and Fermi energy agree well with our asymptotic result. The cancellation of the Ising term and the $z$-th component of the Heisenberg term was first reported in ~\cite{chang2015} for $Q=1$ and we confirm that it is a shared feature for all monopole charges with vertical impurity alignment.

For the intrinsic case, the RKKY interaction becomes nonoscillatory,
\bea
H_{\mathrm{RKKY}}^{z,0} & \propto & \frac{1}{\tz^{4/Q+1}} (\bS_1 \cdot \bS_2 - \beta_1 S_1^z S_2^z),\label{intrinsicz}
\eea
with $\beta_1$ being a real constant to be determined numerically for given $Q$. With the increase of monopole charge, the RKKY interaction becomes more
long-ranged.


\subsection{IMPURITIES IN THE TRANSVERSE PLANE}\label{sinplane}


For magnetic impurities aligned in the transverse plane, $i.e.$, $\tz=0$, the $\tg_i$ coefficients can be simplified to,
\bea
\tg_0 &=&  \int_0^{+\infty}dq \int_{-\pi/2}^{\pi/2}d\phi \frac{\tilde \epsilon q^{2/Q}   (\cos \phi)^{2/Q-1}  J_0\left( (q \cos\phi)^{1/Q} \tilde \rho \right)           }{q^2-\tilde \epsilon^2},\nonumber\\
\tg_1 &=&  i^Q      \int_0^{+\infty}dq \int_{-\pi/2}^{\pi/2}d\phi \frac{ q^{2/Q+1} (\cos \phi)^{2/Q}  J_Q\left( (q \cos\phi)^{1/Q} \tilde \rho \right)  }{q^2-\tilde \epsilon^2},\nonumber\\
\tg_3 &=&  0.\label{gx}
\eea

Analogous heuristic argument can be applied to the above equations.
The coefficients $\tG_{ij}$ take the following form for finite Fermi energy and at long distance,
\begin{equation}
    \tG_{00}(\to_F,\tr)\approx\tG_{11}(\to_F,\tr)\approx \frac{\alpha_2 \to_F^{1/Q+1}}{\tr^{3}}\cos(2\to_F^{1/Q}\tr),\label{G00x}
\end{equation}
where $\alpha_2 = -Q^2 \pi^2/2$ is a real constant determined by monopole charge $Q$.

The asymptotic RKKY interaction for impurities in the transverse plane thus reduces to,
\bea
H_{\mathrm{RKKY}}^\rho & \propto & \frac{\to_F^{1/Q+1}\cos(2\to_F^{1/Q}\tr)}{\tr^{3}} \times sc,\label{hr}
\eea
with the spin correlator

\begin{eqnarray}
    sc=
\left\{\begin{array}{cc}
        \bS_1 \cdot \bS_2 - (\bS_1\cdot\hs) (\bS_2\cdot\hs), & Q\text{ odd,} \\
        (\bS_1\cdot\hs) (\bS_2\cdot\hs),& Q\text{ even.}
                    \end{array}\right.\label{evenodd}
\end{eqnarray}
For odd Q, the spin correlator term is much similar to the that derived in Eq.~(\ref{hz}), only that the preferred spin alignment $\hs$ generally does not coincide with impurity alignment $\hat{\rho}$ when $Q \neq 1$. For even $Q$, the spin-frustrated term dominates the RKKY interaction since the components of the Heisenberg term cancel out.

\begin{figure}
\includegraphics[width=7cm, height=11cm]{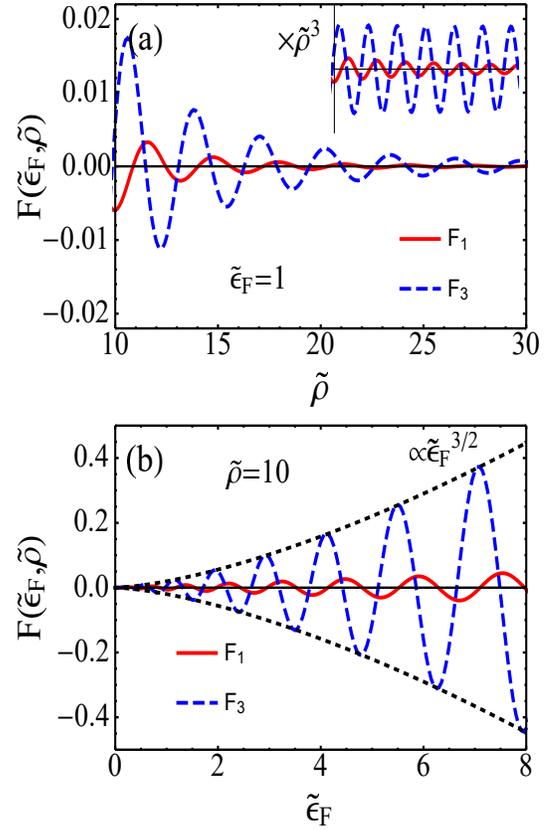}
\caption{The exact RKKY range functions (in units of $J^2 C$) for impurities in the transverse plane. Here $Q=2$. The top and bottom panels show power-law dependence of RKKY range functions on impurity separation and on Fermi energy, respectively. The spin-frustrated term dominates over the Heisenberg term for even monopole charges.} \label{rplot}
\end{figure}

Fig.~\ref{rplot} shows exact numerical results of RKKY range functions in double-Weyl semimetal ($Q=2$) for impurities in the transverse plane, the power-law dependence on impurity separation and Fermi energy agree well with our analytical result. Note that the cancellation of $\tG_{00}$ and $\tG_{11}$ is valid up to the leading order and leaves out a minor Heisenberg term proportional to $\frac{\to_F}{\tr^{4}}\sin(2\to_F^{1/Q}\tr)$ (see solid red lines in Fig.~\ref{rplot}). Similar higher-order residual term is also present between the cancellation of $F_1$ and $F_4$ in previous subsection (not explicitly plotted in Fig.~\ref{zplot}).

For the intrinsic case, the regulated result of the divergent integration $\int_{-\infty}^0 d \to \tg_i^2$ gives
\begin{equation}
    \tG_{00}(0,\tr)\sim\tG_{11}(0,\tr)\propto\frac{1}{\tr^{Q+4}},
\end{equation}
and the RKKY interaction becomes non-oscillatory,
\bea
H_{\mathrm{RKKY}}^{\rho,0} & \propto & \frac{1}{\tr^{Q+4}} \left[\bS_1 \cdot \bS_2 - \beta_2(\bS_1\cdot\hs) (\bS_2\cdot\hs)\right],\label{intrinsicr}
\eea
with $\beta_2$ being a real constant for given $Q$. For large monopole charge, the interaction becomes quite short-ranged
in planar directions.


\subsection{DISCUSSION OF THE RESULTS}

Eq.~(\ref{hz}), Eq.(\ref{intrinsicz}), Eq.(\ref{hr}) and Eq.(\ref{intrinsicr})
constitute the central results of our present work. It is quick to verify that
for $Q=1$ WSMs, the RKKY interaction takes the form of
\begin{equation}
    H_{\mathrm{RKKY}} \propto \frac{\to_F^{2}\cos(2\to_F \tR)}{\tR^{3}} (\bS_1 \cdot \bS_2 - S_1^j S_2^j),
\end{equation}
for finite doping, and 
\begin{equation}
    H_{\mathrm{RKKY}}^{0} \propto \frac{1}{\tR^{5}} (\bS_1 \cdot \bS_2 - \beta S_1^j S_2^j),
\end{equation}
for the intrinsic case, where $j$ denotes the direction of impurity alignment. Both formulas coincide with the results previously obtained in refs.~\cite{chang2015,hosseini2015}.

Compared to the $Q=1$ case, several distinctive features arise for $Q\geq2$. First, for finite doping the RKKY interaction decays as $1/{\tz^{2/Q+1}}$ and $1/{\tr^{3}}$ for two representative impurity alignments (as $1/\tz^{4/Q+1}$ and $1/\tr^{Q+4}$ for the intrinsic case, respectively). With the increase of monopole charge $Q$, it is readily seen that the range functions become more long-ranged in the vertical direction. 
Second, the power-law dependence on the Fermi energy amounts
to $\to_F^{2/Q}$ for the vertical direction and $\to_F^{1/Q+1}$ for the transverse directions, suggesting that
the increase of Fermi energy contributes a bigger boost to the RKKY range functions in
the latter case. Besides, there exists an even-odd discrepancy for impurities in the transverse plane: 
for odd $Q$, the Heisenberg term coexists with the spin-frustrated term; while for even $Q$
the Heisenberg term cancels out up to leading order and 
the spin-frustrated term dominates the RKKY interaction (see Eq.(\ref{evenodd})).



\section{CONCLUSION}\label{concl}

In summary, we have studied the RKKY interactions for
WSMs with arbitrary monopole charge $Q$ and analytically obtained 
their asymptotic expressions in the long range limit. 
The results indicate that the power-law dependence of the RKKY interaction on impurity
separation and Fermi energy is directly controlled by the monopole charge.
As the power-law dependence is tightly related to the monopole charge, it thus provides a potential way
to determine the monopole charge of WSMs. More importantly,
the interaction becomes quite long-ranged and quasi-one-dimensional for WSMs with large monopole charge, which may 
trigger interesting magnetic orders and result in applications in spintronics.

Besides the material candidates predicated by first-principle calculations~\cite{xu2011,huang2016}, recently, several other works postulate that multi-WSMs can be dynamically created either from normal insulators~\cite{gupta2017}
or from crossing nodal line semimetals~\cite{Ezawa2017fmwsm,Yan2017fmwsm}. Considering the rapid development of this field, it is expected that
our theoretical predictions can be experimentally tested in near future.

\section{Acknowledgments}

We are grateful to Zhongbo Yan and Haoran Chang for their helpful discussions.
This work is supported by NSFC under Grant NO. 11375168.

\bibliography{luke}

\begin{thebibliography}{43}%
\makeatletter
\providecommand \@ifxundefined [1]{%
 \@ifx{#1\undefined}
}%
\providecommand \@ifnum [1]{%
 \ifnum #1\expandafter \@firstoftwo
 \else \expandafter \@secondoftwo
 \fi
}%
\providecommand \@ifx [1]{%
 \ifx #1\expandafter \@firstoftwo
 \else \expandafter \@secondoftwo
 \fi
}%
\providecommand \natexlab [1]{#1}%
\providecommand \enquote  [1]{``#1''}%
\providecommand \bibnamefont  [1]{#1}%
\providecommand \bibfnamefont [1]{#1}%
\providecommand \citenamefont [1]{#1}%
\providecommand \href@noop [0]{\@secondoftwo}%
\providecommand \href [0]{\begingroup \@sanitize@url \@href}%
\providecommand \@href[1]{\@@startlink{#1}\@@href}%
\providecommand \@@href[1]{\endgroup#1\@@endlink}%
\providecommand \@sanitize@url [0]{\catcode `\\12\catcode `\$12\catcode
  `\&12\catcode `\#12\catcode `\^12\catcode `\_12\catcode `\%12\relax}%
\providecommand \@@startlink[1]{}%
\providecommand \@@endlink[0]{}%
\providecommand \url  [0]{\begingroup\@sanitize@url \@url }%
\providecommand \@url [1]{\endgroup\@href {#1}{\urlprefix }}%
\providecommand \urlprefix  [0]{URL }%
\providecommand \Eprint [0]{\href }%
\providecommand \doibase [0]{http://dx.doi.org/}%
\providecommand \selectlanguage [0]{\@gobble}%
\providecommand \bibinfo  [0]{\@secondoftwo}%
\providecommand \bibfield  [0]{\@secondoftwo}%
\providecommand \translation [1]{[#1]}%
\providecommand \BibitemOpen [0]{}%
\providecommand \bibitemStop [0]{}%
\providecommand \bibitemNoStop [0]{.\EOS\space}%
\providecommand \EOS [0]{\spacefactor3000\relax}%
\providecommand \BibitemShut  [1]{\csname bibitem#1\endcsname}%
\let\auto@bib@innerbib\@empty
\bibitem [{\citenamefont {Kittel}(2005)}]{kittel2005introduction}%
  \BibitemOpen
  \bibfield  {author} {\bibinfo {author} {\bibfnamefont {Charles}\ \bibnamefont
  {Kittel}},\ }\href@noop {} {\emph {\bibinfo {title} {Introduction to solid
  state physics}}}\ (\bibinfo  {publisher} {Wiley},\ \bibinfo {year}
  {2005})\BibitemShut {NoStop}%
\bibitem [{\citenamefont {Murakami}(2007)}]{Murakami2007weyl}%
  \BibitemOpen
  \bibfield  {author} {\bibinfo {author} {\bibfnamefont {Shuichi}\ \bibnamefont
  {Murakami}},\ }\bibfield  {title} {\enquote {\bibinfo {title} {Phase
  transition between the quantum spin hall and insulator phases in 3d:
  emergence of a topological gapless phase},}\ }\href
  {http://stacks.iop.org/1367-2630/9/i=9/a=356} {\bibfield  {journal} {\bibinfo
   {journal} {New Journal of Physics}\ }\textbf {\bibinfo {volume} {9}},\
  \bibinfo {pages} {356} (\bibinfo {year} {2007})}\BibitemShut {NoStop}%
\bibitem [{\citenamefont {Wan}\ \emph {et~al.}(2011)\citenamefont {Wan},
  \citenamefont {Turner}, \citenamefont {Vishwanath},\ and\ \citenamefont
  {Savrasov}}]{wan2011weyl}%
  \BibitemOpen
  \bibfield  {author} {\bibinfo {author} {\bibfnamefont {Xiangang}\
  \bibnamefont {Wan}}, \bibinfo {author} {\bibfnamefont {Ari~M.}\ \bibnamefont
  {Turner}}, \bibinfo {author} {\bibfnamefont {Ashvin}\ \bibnamefont
  {Vishwanath}}, \ and\ \bibinfo {author} {\bibfnamefont {Sergey~Y.}\
  \bibnamefont {Savrasov}},\ }\bibfield  {title} {\enquote {\bibinfo {title}
  {Topological semimetal and fermi-arc surface states in the electronic
  structure of pyrochlore iridates},}\ }\href {\doibase
  10.1103/PhysRevB.83.205101} {\bibfield  {journal} {\bibinfo  {journal} {Phys.
  Rev. B}\ }\textbf {\bibinfo {volume} {83}},\ \bibinfo {pages} {205101}
  (\bibinfo {year} {2011})}\BibitemShut {NoStop}%
\bibitem [{\citenamefont {Burkov}\ and\ \citenamefont
  {Balents}(2011)}]{Burkov2011weyl}%
  \BibitemOpen
  \bibfield  {author} {\bibinfo {author} {\bibfnamefont {A.~A.}\ \bibnamefont
  {Burkov}}\ and\ \bibinfo {author} {\bibfnamefont {Leon}\ \bibnamefont
  {Balents}},\ }\bibfield  {title} {\enquote {\bibinfo {title} {Weyl semimetal
  in a topological insulator multilayer},}\ }\href {\doibase
  10.1103/PhysRevLett.107.127205} {\bibfield  {journal} {\bibinfo  {journal}
  {Phys. Rev. Lett.}\ }\textbf {\bibinfo {volume} {107}},\ \bibinfo {pages}
  {127205} (\bibinfo {year} {2011})}\BibitemShut {NoStop}%
\bibitem [{\citenamefont {Weng}\ \emph {et~al.}(2015)\citenamefont {Weng},
  \citenamefont {Fang}, \citenamefont {Fang}, \citenamefont {Bernevig},\ and\
  \citenamefont {Dai}}]{Weng2015weyl}%
  \BibitemOpen
  \bibfield  {author} {\bibinfo {author} {\bibfnamefont {Hongming}\
  \bibnamefont {Weng}}, \bibinfo {author} {\bibfnamefont {Chen}\ \bibnamefont
  {Fang}}, \bibinfo {author} {\bibfnamefont {Zhong}\ \bibnamefont {Fang}},
  \bibinfo {author} {\bibfnamefont {B.~Andrei}\ \bibnamefont {Bernevig}}, \
  and\ \bibinfo {author} {\bibfnamefont {Xi}~\bibnamefont {Dai}},\ }\bibfield
  {title} {\enquote {\bibinfo {title} {Weyl semimetal phase in
  noncentrosymmetric transition-metal monophosphides},}\ }\href {\doibase
  10.1103/PhysRevX.5.011029} {\bibfield  {journal} {\bibinfo  {journal} {Phys.
  Rev. X}\ }\textbf {\bibinfo {volume} {5}},\ \bibinfo {pages} {011029}
  (\bibinfo {year} {2015})}\BibitemShut {NoStop}%
\bibitem [{\citenamefont {Huang}\ \emph
  {et~al.}(2015{\natexlab{a}})\citenamefont {Huang}, \citenamefont {Xu},
  \citenamefont {Belopolski}, \citenamefont {Lee}, \citenamefont {Chang},
  \citenamefont {Wang}, \citenamefont {Alidoust}, \citenamefont {Bian},
  \citenamefont {Neupane}, \citenamefont {Zhang} \emph
  {et~al.}}]{huang2015weyl}%
  \BibitemOpen
  \bibfield  {author} {\bibinfo {author} {\bibfnamefont {Shin-Ming}\
  \bibnamefont {Huang}}, \bibinfo {author} {\bibfnamefont {Su-Yang}\
  \bibnamefont {Xu}}, \bibinfo {author} {\bibfnamefont {Ilya}\ \bibnamefont
  {Belopolski}}, \bibinfo {author} {\bibfnamefont {Chi-Cheng}\ \bibnamefont
  {Lee}}, \bibinfo {author} {\bibfnamefont {Guoqing}\ \bibnamefont {Chang}},
  \bibinfo {author} {\bibfnamefont {BaoKai}\ \bibnamefont {Wang}}, \bibinfo
  {author} {\bibfnamefont {Nasser}\ \bibnamefont {Alidoust}}, \bibinfo {author}
  {\bibfnamefont {Guang}\ \bibnamefont {Bian}}, \bibinfo {author}
  {\bibfnamefont {Madhab}\ \bibnamefont {Neupane}}, \bibinfo {author}
  {\bibfnamefont {Chenglong}\ \bibnamefont {Zhang}},  \emph {et~al.},\
  }\bibfield  {title} {\enquote {\bibinfo {title} {A weyl fermion semimetal
  with surface fermi arcs in the transition metal monopnictide taas class},}\
  }\href@noop {} {\bibfield  {journal} {\bibinfo  {journal} {Nature
  communications}\ }\textbf {\bibinfo {volume} {6}},\ \bibinfo {pages} {7373}
  (\bibinfo {year} {2015}{\natexlab{a}})}\BibitemShut {NoStop}%
\bibitem [{\citenamefont {Xu}\ \emph {et~al.}(2015)\citenamefont {Xu},
  \citenamefont {Belopolski}, \citenamefont {Alidoust}, \citenamefont
  {Neupane}, \citenamefont {Bian}, \citenamefont {Zhang}, \citenamefont
  {Sankar}, \citenamefont {Chang}, \citenamefont {Yuan}, \citenamefont {Lee}
  \emph {et~al.}}]{Xu2015weyl}%
  \BibitemOpen
  \bibfield  {author} {\bibinfo {author} {\bibfnamefont {Su-Yang}\ \bibnamefont
  {Xu}}, \bibinfo {author} {\bibfnamefont {Ilya}\ \bibnamefont {Belopolski}},
  \bibinfo {author} {\bibfnamefont {Nasser}\ \bibnamefont {Alidoust}}, \bibinfo
  {author} {\bibfnamefont {Madhab}\ \bibnamefont {Neupane}}, \bibinfo {author}
  {\bibfnamefont {Guang}\ \bibnamefont {Bian}}, \bibinfo {author}
  {\bibfnamefont {Chenglong}\ \bibnamefont {Zhang}}, \bibinfo {author}
  {\bibfnamefont {Raman}\ \bibnamefont {Sankar}}, \bibinfo {author}
  {\bibfnamefont {Guoqing}\ \bibnamefont {Chang}}, \bibinfo {author}
  {\bibfnamefont {Zhujun}\ \bibnamefont {Yuan}}, \bibinfo {author}
  {\bibfnamefont {Chi-Cheng}\ \bibnamefont {Lee}},  \emph {et~al.},\ }\bibfield
   {title} {\enquote {\bibinfo {title} {Discovery of a weyl fermion semimetal
  and topological fermi arcs},}\ }\href@noop {} {\bibfield  {journal} {\bibinfo
   {journal} {Science}\ }\textbf {\bibinfo {volume} {349}},\ \bibinfo {pages}
  {613--617} (\bibinfo {year} {2015})}\BibitemShut {NoStop}%
\bibitem [{\citenamefont {Lv}\ \emph {et~al.}(2015)\citenamefont {Lv},
  \citenamefont {Weng}, \citenamefont {Fu}, \citenamefont {Wang}, \citenamefont
  {Miao}, \citenamefont {Ma}, \citenamefont {Richard}, \citenamefont {Huang},
  \citenamefont {Zhao}, \citenamefont {Chen}, \citenamefont {Fang},
  \citenamefont {Dai}, \citenamefont {Qian},\ and\ \citenamefont
  {Ding}}]{Lv2015weyl}%
  \BibitemOpen
  \bibfield  {author} {\bibinfo {author} {\bibfnamefont {B.~Q.}\ \bibnamefont
  {Lv}}, \bibinfo {author} {\bibfnamefont {H.~M.}\ \bibnamefont {Weng}},
  \bibinfo {author} {\bibfnamefont {B.~B.}\ \bibnamefont {Fu}}, \bibinfo
  {author} {\bibfnamefont {X.~P.}\ \bibnamefont {Wang}}, \bibinfo {author}
  {\bibfnamefont {H.}~\bibnamefont {Miao}}, \bibinfo {author} {\bibfnamefont
  {J.}~\bibnamefont {Ma}}, \bibinfo {author} {\bibfnamefont {P.}~\bibnamefont
  {Richard}}, \bibinfo {author} {\bibfnamefont {X.~C.}\ \bibnamefont {Huang}},
  \bibinfo {author} {\bibfnamefont {L.~X.}\ \bibnamefont {Zhao}}, \bibinfo
  {author} {\bibfnamefont {G.~F.}\ \bibnamefont {Chen}}, \bibinfo {author}
  {\bibfnamefont {Z.}~\bibnamefont {Fang}}, \bibinfo {author} {\bibfnamefont
  {X.}~\bibnamefont {Dai}}, \bibinfo {author} {\bibfnamefont {T.}~\bibnamefont
  {Qian}}, \ and\ \bibinfo {author} {\bibfnamefont {H.}~\bibnamefont {Ding}},\
  }\bibfield  {title} {\enquote {\bibinfo {title} {Experimental discovery of
  weyl semimetal taas},}\ }\href {\doibase 10.1103/PhysRevX.5.031013}
  {\bibfield  {journal} {\bibinfo  {journal} {Phys. Rev. X}\ }\textbf {\bibinfo
  {volume} {5}},\ \bibinfo {pages} {031013} (\bibinfo {year}
  {2015})}\BibitemShut {NoStop}%
\bibitem [{\citenamefont {Lu}\ \emph {et~al.}(2015)\citenamefont {Lu},
  \citenamefont {Wang}, \citenamefont {Ye}, \citenamefont {Ran}, \citenamefont
  {Fu}, \citenamefont {Joannopoulos},\ and\ \citenamefont
  {Solja{\v{c}}i{\'c}}}]{lu2015weyl}%
  \BibitemOpen
  \bibfield  {author} {\bibinfo {author} {\bibfnamefont {Ling}\ \bibnamefont
  {Lu}}, \bibinfo {author} {\bibfnamefont {Zhiyu}\ \bibnamefont {Wang}},
  \bibinfo {author} {\bibfnamefont {Dexin}\ \bibnamefont {Ye}}, \bibinfo
  {author} {\bibfnamefont {Lixin}\ \bibnamefont {Ran}}, \bibinfo {author}
  {\bibfnamefont {Liang}\ \bibnamefont {Fu}}, \bibinfo {author} {\bibfnamefont
  {John~D}\ \bibnamefont {Joannopoulos}}, \ and\ \bibinfo {author}
  {\bibfnamefont {Marin}\ \bibnamefont {Solja{\v{c}}i{\'c}}},\ }\bibfield
  {title} {\enquote {\bibinfo {title} {Experimental observation of weyl
  points},}\ }\href@noop {} {\bibfield  {journal} {\bibinfo  {journal}
  {Science}\ }\textbf {\bibinfo {volume} {349}},\ \bibinfo {pages} {622--624}
  (\bibinfo {year} {2015})}\BibitemShut {NoStop}%
\bibitem [{\citenamefont {Xiao}\ \emph {et~al.}(2010)\citenamefont {Xiao},
  \citenamefont {Chang},\ and\ \citenamefont {Niu}}]{Xiao2010berry}%
  \BibitemOpen
  \bibfield  {author} {\bibinfo {author} {\bibfnamefont {Di}~\bibnamefont
  {Xiao}}, \bibinfo {author} {\bibfnamefont {Ming-Che}\ \bibnamefont {Chang}},
  \ and\ \bibinfo {author} {\bibfnamefont {Qian}\ \bibnamefont {Niu}},\
  }\bibfield  {title} {\enquote {\bibinfo {title} {Berry phase effects on
  electronic properties},}\ }\href {\doibase 10.1103/RevModPhys.82.1959}
  {\bibfield  {journal} {\bibinfo  {journal} {Rev. Mod. Phys.}\ }\textbf
  {\bibinfo {volume} {82}},\ \bibinfo {pages} {1959--2007} (\bibinfo {year}
  {2010})}\BibitemShut {NoStop}%
\bibitem [{\citenamefont {Nielsen}\ and\ \citenamefont
  {Ninomiya}(1983)}]{nielsen1983adler}%
  \BibitemOpen
  \bibfield  {author} {\bibinfo {author} {\bibfnamefont {Holger~Bech}\
  \bibnamefont {Nielsen}}\ and\ \bibinfo {author} {\bibfnamefont {Masao}\
  \bibnamefont {Ninomiya}},\ }\bibfield  {title} {\enquote {\bibinfo {title}
  {The adler-bell-jackiw anomaly and weyl fermions in a crystal},}\ }\href@noop
  {} {\bibfield  {journal} {\bibinfo  {journal} {Physics Letters B}\ }\textbf
  {\bibinfo {volume} {130}},\ \bibinfo {pages} {389--396} (\bibinfo {year}
  {1983})}\BibitemShut {NoStop}%
\bibitem [{\citenamefont {Son}\ and\ \citenamefont
  {Spivak}(2013)}]{Son2013anomaly}%
  \BibitemOpen
  \bibfield  {author} {\bibinfo {author} {\bibfnamefont {D.~T.}\ \bibnamefont
  {Son}}\ and\ \bibinfo {author} {\bibfnamefont {B.~Z.}\ \bibnamefont
  {Spivak}},\ }\bibfield  {title} {\enquote {\bibinfo {title} {Chiral anomaly
  and classical negative magnetoresistance of weyl metals},}\ }\href {\doibase
  10.1103/PhysRevB.88.104412} {\bibfield  {journal} {\bibinfo  {journal} {Phys.
  Rev. B}\ }\textbf {\bibinfo {volume} {88}},\ \bibinfo {pages} {104412}
  (\bibinfo {year} {2013})}\BibitemShut {NoStop}%
\bibitem [{\citenamefont {Huang}\ \emph
  {et~al.}(2015{\natexlab{b}})\citenamefont {Huang}, \citenamefont {Zhao},
  \citenamefont {Long}, \citenamefont {Wang}, \citenamefont {Chen},
  \citenamefont {Yang}, \citenamefont {Liang}, \citenamefont {Xue},
  \citenamefont {Weng}, \citenamefont {Fang}, \citenamefont {Dai},\ and\
  \citenamefont {Chen}}]{Huang2015anomaly}%
  \BibitemOpen
  \bibfield  {author} {\bibinfo {author} {\bibfnamefont {Xiaochun}\
  \bibnamefont {Huang}}, \bibinfo {author} {\bibfnamefont {Lingxiao}\
  \bibnamefont {Zhao}}, \bibinfo {author} {\bibfnamefont {Yujia}\ \bibnamefont
  {Long}}, \bibinfo {author} {\bibfnamefont {Peipei}\ \bibnamefont {Wang}},
  \bibinfo {author} {\bibfnamefont {Dong}\ \bibnamefont {Chen}}, \bibinfo
  {author} {\bibfnamefont {Zhanhai}\ \bibnamefont {Yang}}, \bibinfo {author}
  {\bibfnamefont {Hui}\ \bibnamefont {Liang}}, \bibinfo {author} {\bibfnamefont
  {Mianqi}\ \bibnamefont {Xue}}, \bibinfo {author} {\bibfnamefont {Hongming}\
  \bibnamefont {Weng}}, \bibinfo {author} {\bibfnamefont {Zhong}\ \bibnamefont
  {Fang}}, \bibinfo {author} {\bibfnamefont {Xi}~\bibnamefont {Dai}}, \ and\
  \bibinfo {author} {\bibfnamefont {Genfu}\ \bibnamefont {Chen}},\ }\bibfield
  {title} {\enquote {\bibinfo {title} {Observation of the
  chiral-anomaly-induced negative magnetoresistance in 3d weyl semimetal
  taas},}\ }\href {\doibase 10.1103/PhysRevX.5.031023} {\bibfield  {journal}
  {\bibinfo  {journal} {Phys. Rev. X}\ }\textbf {\bibinfo {volume} {5}},\
  \bibinfo {pages} {031023} (\bibinfo {year} {2015}{\natexlab{b}})}\BibitemShut
  {NoStop}%
\bibitem [{\citenamefont {Parameswaran}\ \emph {et~al.}(2014)\citenamefont
  {Parameswaran}, \citenamefont {Grover}, \citenamefont {Abanin}, \citenamefont
  {Pesin},\ and\ \citenamefont {Vishwanath}}]{Parameswaran2014anomaly}%
  \BibitemOpen
  \bibfield  {author} {\bibinfo {author} {\bibfnamefont {S.~A.}\ \bibnamefont
  {Parameswaran}}, \bibinfo {author} {\bibfnamefont {T.}~\bibnamefont
  {Grover}}, \bibinfo {author} {\bibfnamefont {D.~A.}\ \bibnamefont {Abanin}},
  \bibinfo {author} {\bibfnamefont {D.~A.}\ \bibnamefont {Pesin}}, \ and\
  \bibinfo {author} {\bibfnamefont {A.}~\bibnamefont {Vishwanath}},\ }\bibfield
   {title} {\enquote {\bibinfo {title} {Probing the chiral anomaly with
  nonlocal transport in three-dimensional topological semimetals},}\ }\href
  {\doibase 10.1103/PhysRevX.4.031035} {\bibfield  {journal} {\bibinfo
  {journal} {Phys. Rev. X}\ }\textbf {\bibinfo {volume} {4}},\ \bibinfo {pages}
  {031035} (\bibinfo {year} {2014})}\BibitemShut {NoStop}%
\bibitem [{\citenamefont {Zhang}\ \emph {et~al.}(2016)\citenamefont {Zhang},
  \citenamefont {Xu}, \citenamefont {Belopolski}, \citenamefont {Yuan},
  \citenamefont {Lin}, \citenamefont {Tong}, \citenamefont {Bian},
  \citenamefont {Alidoust}, \citenamefont {Lee}, \citenamefont {Huang} \emph
  {et~al.}}]{Zhang2016anomaly}%
  \BibitemOpen
  \bibfield  {author} {\bibinfo {author} {\bibfnamefont {Cheng-Long}\
  \bibnamefont {Zhang}}, \bibinfo {author} {\bibfnamefont {Su-Yang}\
  \bibnamefont {Xu}}, \bibinfo {author} {\bibfnamefont {Ilya}\ \bibnamefont
  {Belopolski}}, \bibinfo {author} {\bibfnamefont {Zhujun}\ \bibnamefont
  {Yuan}}, \bibinfo {author} {\bibfnamefont {Ziquan}\ \bibnamefont {Lin}},
  \bibinfo {author} {\bibfnamefont {Bingbing}\ \bibnamefont {Tong}}, \bibinfo
  {author} {\bibfnamefont {Guang}\ \bibnamefont {Bian}}, \bibinfo {author}
  {\bibfnamefont {Nasser}\ \bibnamefont {Alidoust}}, \bibinfo {author}
  {\bibfnamefont {Chi-Cheng}\ \bibnamefont {Lee}}, \bibinfo {author}
  {\bibfnamefont {Shin-Ming}\ \bibnamefont {Huang}},  \emph {et~al.},\
  }\bibfield  {title} {\enquote {\bibinfo {title} {Signatures of the
  adler--bell--jackiw chiral anomaly in a weyl fermion semimetal},}\ }\href
  {\doibase 10.1038/ncomms10735} {\bibfield  {journal} {\bibinfo  {journal}
  {Nature communications}\ }\textbf {\bibinfo {volume} {7}} (\bibinfo {year}
  {2016}),\ 10.1038/ncomms10735}\BibitemShut {NoStop}%
\bibitem [{\citenamefont {Yan}\ and\ \citenamefont
  {Felser}(2017)}]{yan2017review}%
  \BibitemOpen
  \bibfield  {author} {\bibinfo {author} {\bibfnamefont {Binghai}\ \bibnamefont
  {Yan}}\ and\ \bibinfo {author} {\bibfnamefont {Claudia}\ \bibnamefont
  {Felser}},\ }\bibfield  {title} {\enquote {\bibinfo {title} {Topological
  materials: Weyl semimetals},}\ }\href {\doibase
  10.1146/annurev-conmatphys-031016-025458} {\bibfield  {journal} {\bibinfo
  {journal} {Annual Review of Condensed Matter Physics}\ }\textbf {\bibinfo
  {volume} {8}},\ \bibinfo {pages} {337--354} (\bibinfo {year}
  {2017})}\BibitemShut {NoStop}%
\bibitem [{\citenamefont {{Armitage}}\ \emph {et~al.}(2017)\citenamefont
  {{Armitage}}, \citenamefont {{Mele}},\ and\ \citenamefont
  {{Vishwanath}}}]{Armitage2017review}%
  \BibitemOpen
  \bibfield  {author} {\bibinfo {author} {\bibfnamefont {N.~P.}\ \bibnamefont
  {{Armitage}}}, \bibinfo {author} {\bibfnamefont {E.~J.}\ \bibnamefont
  {{Mele}}}, \ and\ \bibinfo {author} {\bibfnamefont {A.}~\bibnamefont
  {{Vishwanath}}},\ }\bibfield  {title} {\enquote {\bibinfo {title} {{Weyl and
  Dirac Semimetals in Three Dimensional Solids}},}\ }\href@noop {} {\bibfield
  {journal} {\bibinfo  {journal} {ArXiv e-prints}\ } (\bibinfo {year}
  {2017})},\ \Eprint {http://arxiv.org/abs/1705.01111} {arXiv:1705.01111
  [cond-mat.str-el]} \BibitemShut {NoStop}%
\bibitem [{\citenamefont {{Burkov}}(2017)}]{Burkov2017review}%
  \BibitemOpen
  \bibfield  {author} {\bibinfo {author} {\bibfnamefont {A.~A.}\ \bibnamefont
  {{Burkov}}},\ }\bibfield  {title} {\enquote {\bibinfo {title} {{Weyl
  Metals}},}\ }\href@noop {} {\bibfield  {journal} {\bibinfo  {journal} {ArXiv
  e-prints}\ } (\bibinfo {year} {2017})},\ \Eprint
  {http://arxiv.org/abs/1704.06660} {arXiv:1704.06660 [cond-mat.mes-hall]}
  \BibitemShut {NoStop}%
\bibitem [{\citenamefont {Hasan}\ \emph {et~al.}(2017)\citenamefont {Hasan},
  \citenamefont {Xu}, \citenamefont {Belopolski},\ and\ \citenamefont
  {Huang}}]{Hasan2017review}%
  \BibitemOpen
  \bibfield  {author} {\bibinfo {author} {\bibfnamefont {M.~Zahid}\
  \bibnamefont {Hasan}}, \bibinfo {author} {\bibfnamefont {Su-Yang}\
  \bibnamefont {Xu}}, \bibinfo {author} {\bibfnamefont {Ilya}\ \bibnamefont
  {Belopolski}}, \ and\ \bibinfo {author} {\bibfnamefont {Shin-Ming}\
  \bibnamefont {Huang}},\ }\bibfield  {title} {\enquote {\bibinfo {title}
  {Discovery of weyl fermion semimetals and topological fermi arc states},}\
  }\href {\doibase 10.1146/annurev-conmatphys-031016-025225} {\bibfield
  {journal} {\bibinfo  {journal} {Annual Review of Condensed Matter Physics}\
  }\textbf {\bibinfo {volume} {8}},\ \bibinfo {pages} {289--309} (\bibinfo
  {year} {2017})}\BibitemShut {NoStop}%
\bibitem [{\citenamefont {Xu}\ \emph {et~al.}(2011)\citenamefont {Xu},
  \citenamefont {Weng}, \citenamefont {Wang}, \citenamefont {Dai},\ and\
  \citenamefont {Fang}}]{xu2011}%
  \BibitemOpen
  \bibfield  {author} {\bibinfo {author} {\bibfnamefont {Gang}\ \bibnamefont
  {Xu}}, \bibinfo {author} {\bibfnamefont {Hongming}\ \bibnamefont {Weng}},
  \bibinfo {author} {\bibfnamefont {Zhijun}\ \bibnamefont {Wang}}, \bibinfo
  {author} {\bibfnamefont {Xi}~\bibnamefont {Dai}}, \ and\ \bibinfo {author}
  {\bibfnamefont {Zhong}\ \bibnamefont {Fang}},\ }\bibfield  {title} {\enquote
  {\bibinfo {title} {Chern {{Semimetal}} and the {{Quantized Anomalous Hall
  Effect}} in ${\mathrm{hgcr}}_{2}{\mathrm{se}}_{4}$},}\ }\href {\doibase
  10.1103/PhysRevLett.107.186806} {\bibfield  {journal} {\bibinfo  {journal}
  {Phys. Rev. Lett.}\ }\textbf {\bibinfo {volume} {107}},\ \bibinfo {pages}
  {186806} (\bibinfo {year} {2011})}\BibitemShut {NoStop}%
\bibitem [{\citenamefont {Fang}\ \emph {et~al.}(2012)\citenamefont {Fang},
  \citenamefont {Gilbert}, \citenamefont {Dai},\ and\ \citenamefont
  {Bernevig}}]{Fang2012mwsm}%
  \BibitemOpen
  \bibfield  {author} {\bibinfo {author} {\bibfnamefont {Chen}\ \bibnamefont
  {Fang}}, \bibinfo {author} {\bibfnamefont {Matthew~J.}\ \bibnamefont
  {Gilbert}}, \bibinfo {author} {\bibfnamefont {Xi}~\bibnamefont {Dai}}, \ and\
  \bibinfo {author} {\bibfnamefont {B.~Andrei}\ \bibnamefont {Bernevig}},\
  }\bibfield  {title} {\enquote {\bibinfo {title} {Multi-weyl topological
  semimetals stabilized by point group symmetry},}\ }\href {\doibase
  10.1103/PhysRevLett.108.266802} {\bibfield  {journal} {\bibinfo  {journal}
  {Phys. Rev. Lett.}\ }\textbf {\bibinfo {volume} {108}},\ \bibinfo {pages}
  {266802} (\bibinfo {year} {2012})}\BibitemShut {NoStop}%
\bibitem [{\citenamefont {Huang}\ \emph {et~al.}(2016)\citenamefont {Huang},
  \citenamefont {Xu}, \citenamefont {Belopolski}, \citenamefont {Lee},
  \citenamefont {Chang}, \citenamefont {Chang}, \citenamefont {Wang},
  \citenamefont {Alidoust}, \citenamefont {Bian}, \citenamefont {Neupane},
  \citenamefont {Sanchez}, \citenamefont {Zheng}, \citenamefont {Jeng},
  \citenamefont {Bansil}, \citenamefont {Neupert}, \citenamefont {Lin},\ and\
  \citenamefont {Hasan}}]{huang2016}%
  \BibitemOpen
  \bibfield  {author} {\bibinfo {author} {\bibfnamefont {Shin-Ming}\
  \bibnamefont {Huang}}, \bibinfo {author} {\bibfnamefont {Su-Yang}\
  \bibnamefont {Xu}}, \bibinfo {author} {\bibfnamefont {Ilya}\ \bibnamefont
  {Belopolski}}, \bibinfo {author} {\bibfnamefont {Chi-Cheng}\ \bibnamefont
  {Lee}}, \bibinfo {author} {\bibfnamefont {Guoqing}\ \bibnamefont {Chang}},
  \bibinfo {author} {\bibfnamefont {Tay-Rong}\ \bibnamefont {Chang}}, \bibinfo
  {author} {\bibfnamefont {BaoKai}\ \bibnamefont {Wang}}, \bibinfo {author}
  {\bibfnamefont {Nasser}\ \bibnamefont {Alidoust}}, \bibinfo {author}
  {\bibfnamefont {Guang}\ \bibnamefont {Bian}}, \bibinfo {author}
  {\bibfnamefont {Madhab}\ \bibnamefont {Neupane}}, \bibinfo {author}
  {\bibfnamefont {Daniel}\ \bibnamefont {Sanchez}}, \bibinfo {author}
  {\bibfnamefont {Hao}\ \bibnamefont {Zheng}}, \bibinfo {author} {\bibfnamefont
  {Horng-Tay}\ \bibnamefont {Jeng}}, \bibinfo {author} {\bibfnamefont {Arun}\
  \bibnamefont {Bansil}}, \bibinfo {author} {\bibfnamefont {Titus}\
  \bibnamefont {Neupert}}, \bibinfo {author} {\bibfnamefont {Hsin}\
  \bibnamefont {Lin}}, \ and\ \bibinfo {author} {\bibfnamefont {M.~Zahid}\
  \bibnamefont {Hasan}},\ }\bibfield  {title} {\enquote {\bibinfo {title} {New
  type of {{Weyl}} semimetal with quadratic double {{Weyl}} fermions},}\ }\href
  {\doibase 10.1073/pnas.1514581113} {\bibfield  {journal} {\bibinfo  {journal}
  {Proc. Nat. Acad. Sci. USA}\ }\textbf {\bibinfo {volume} {113}},\ \bibinfo
  {pages} {1180--1185} (\bibinfo {year} {2016})}\BibitemShut {NoStop}%
\bibitem [{\citenamefont {Lai}(2015)}]{Lai2015dwsm}%
  \BibitemOpen
  \bibfield  {author} {\bibinfo {author} {\bibfnamefont {Hsin-Hua}\
  \bibnamefont {Lai}},\ }\bibfield  {title} {\enquote {\bibinfo {title}
  {Correlation effects in double-weyl semimetals},}\ }\href {\doibase
  10.1103/PhysRevB.91.235131} {\bibfield  {journal} {\bibinfo  {journal} {Phys.
  Rev. B}\ }\textbf {\bibinfo {volume} {91}},\ \bibinfo {pages} {235131}
  (\bibinfo {year} {2015})}\BibitemShut {NoStop}%
\bibitem [{\citenamefont {Jian}\ and\ \citenamefont
  {Yao}(2015)}]{Jian2015dwsm}%
  \BibitemOpen
  \bibfield  {author} {\bibinfo {author} {\bibfnamefont {Shao-Kai}\
  \bibnamefont {Jian}}\ and\ \bibinfo {author} {\bibfnamefont {Hong}\
  \bibnamefont {Yao}},\ }\bibfield  {title} {\enquote {\bibinfo {title}
  {Correlated double-weyl semimetals with coulomb interactions: Possible
  applications to ${\mathrm{hgcr}}_{2}{\mathrm{se}}_{4}$ and
  ${\mathrm{srsi}}_{2}$},}\ }\href {\doibase 10.1103/PhysRevB.92.045121}
  {\bibfield  {journal} {\bibinfo  {journal} {Phys. Rev. B}\ }\textbf {\bibinfo
  {volume} {92}},\ \bibinfo {pages} {045121} (\bibinfo {year}
  {2015})}\BibitemShut {NoStop}%
\bibitem [{\citenamefont {Ahn}\ \emph {et~al.}(2016)\citenamefont {Ahn},
  \citenamefont {Hwang},\ and\ \citenamefont {Min}}]{Ahn2016mwsm}%
  \BibitemOpen
  \bibfield  {author} {\bibinfo {author} {\bibfnamefont {Seongjin}\
  \bibnamefont {Ahn}}, \bibinfo {author} {\bibfnamefont {EH}~\bibnamefont
  {Hwang}}, \ and\ \bibinfo {author} {\bibfnamefont {Hongki}\ \bibnamefont
  {Min}},\ }\bibfield  {title} {\enquote {\bibinfo {title} {Collective modes in
  multi-weyl semimetals},}\ }\href {\doibase 10.1038/srep34023} {\bibfield
  {journal} {\bibinfo  {journal} {Scientific reports}\ }\textbf {\bibinfo
  {volume} {6}},\ \bibinfo {pages} {34023} (\bibinfo {year}
  {2016})}\BibitemShut {NoStop}%
\bibitem [{\citenamefont {Ahn}\ \emph {et~al.}(2017)\citenamefont {Ahn},
  \citenamefont {Mele},\ and\ \citenamefont {Min}}]{Ahn2017mwsm}%
  \BibitemOpen
  \bibfield  {author} {\bibinfo {author} {\bibfnamefont {Seongjin}\
  \bibnamefont {Ahn}}, \bibinfo {author} {\bibfnamefont {E.~J.}\ \bibnamefont
  {Mele}}, \ and\ \bibinfo {author} {\bibfnamefont {Hongki}\ \bibnamefont
  {Min}},\ }\bibfield  {title} {\enquote {\bibinfo {title} {Optical
  conductivity of multi-weyl semimetals},}\ }\href {\doibase
  10.1103/PhysRevB.95.161112} {\bibfield  {journal} {\bibinfo  {journal} {Phys.
  Rev. B}\ }\textbf {\bibinfo {volume} {95}},\ \bibinfo {pages} {161112}
  (\bibinfo {year} {2017})}\BibitemShut {NoStop}%
\bibitem [{\citenamefont {Park}\ \emph {et~al.}(2017)\citenamefont {Park},
  \citenamefont {Woo}, \citenamefont {Mele},\ and\ \citenamefont
  {Min}}]{Park2017mwsm}%
  \BibitemOpen
  \bibfield  {author} {\bibinfo {author} {\bibfnamefont {Sanghyun}\
  \bibnamefont {Park}}, \bibinfo {author} {\bibfnamefont {Seungchan}\
  \bibnamefont {Woo}}, \bibinfo {author} {\bibfnamefont {E.~J.}\ \bibnamefont
  {Mele}}, \ and\ \bibinfo {author} {\bibfnamefont {Hongki}\ \bibnamefont
  {Min}},\ }\bibfield  {title} {\enquote {\bibinfo {title} {Semiclassical
  boltzmann transport theory for multi-weyl semimetals},}\ }\href {\doibase
  10.1103/PhysRevB.95.161113} {\bibfield  {journal} {\bibinfo  {journal} {Phys.
  Rev. B}\ }\textbf {\bibinfo {volume} {95}},\ \bibinfo {pages} {161113}
  (\bibinfo {year} {2017})}\BibitemShut {NoStop}%
\bibitem [{\citenamefont {{Zhang}}\ \emph {et~al.}(2016)\citenamefont
  {{Zhang}}, \citenamefont {{Jian}},\ and\ \citenamefont
  {{Yao}}}]{Zhang2016mwsm}%
  \BibitemOpen
  \bibfield  {author} {\bibinfo {author} {\bibfnamefont {S.-X.}\ \bibnamefont
  {{Zhang}}}, \bibinfo {author} {\bibfnamefont {S.-K.}\ \bibnamefont {{Jian}}},
  \ and\ \bibinfo {author} {\bibfnamefont {H.}~\bibnamefont {{Yao}}},\
  }\bibfield  {title} {\enquote {\bibinfo {title} {{Correlated triple-Weyl
  semimetals with Coulomb interactions}},}\ }\href@noop {} {\bibfield
  {journal} {\bibinfo  {journal} {ArXiv e-prints}\ } (\bibinfo {year}
  {2016})},\ \Eprint {http://arxiv.org/abs/1610.08975} {arXiv:1610.08975
  [cond-mat.str-el]} \BibitemShut {NoStop}%
\bibitem [{\citenamefont {{Wang}}\ \emph {et~al.}(2017)\citenamefont {{Wang}},
  \citenamefont {{Liu}},\ and\ \citenamefont {{Zhang}}}]{Wang2017mwsm}%
  \BibitemOpen
  \bibfield  {author} {\bibinfo {author} {\bibfnamefont {J.-R.}\ \bibnamefont
  {{Wang}}}, \bibinfo {author} {\bibfnamefont {G.-Z.}\ \bibnamefont {{Liu}}}, \
  and\ \bibinfo {author} {\bibfnamefont {C.-J.}\ \bibnamefont {{Zhang}}},\
  }\bibfield  {title} {\enquote {\bibinfo {title} {{Quantum phase transition
  and non-Fermi liquid behavior in multi-Weyl semimetals}},}\ }\href@noop {}
  {\bibfield  {journal} {\bibinfo  {journal} {ArXiv e-prints}\ } (\bibinfo
  {year} {2017})},\ \Eprint {http://arxiv.org/abs/1705.04001} {arXiv:1705.04001
  [cond-mat.str-el]} \BibitemShut {NoStop}%
\bibitem [{\citenamefont {{Sun}}\ and\ \citenamefont
  {{Wang}}(2017)}]{Sun2017dwsm}%
  \BibitemOpen
  \bibfield  {author} {\bibinfo {author} {\bibfnamefont {Y.}~\bibnamefont
  {{Sun}}}\ and\ \bibinfo {author} {\bibfnamefont {A.-M.}\ \bibnamefont
  {{Wang}}},\ }\bibfield  {title} {\enquote {\bibinfo {title} {{Magneto-optical
  conductivity of double-Weyl semimetals}},}\ }\href@noop {} {\bibfield
  {journal} {\bibinfo  {journal} {ArXiv e-prints}\ } (\bibinfo {year}
  {2017})},\ \Eprint {http://arxiv.org/abs/1705.02695} {arXiv:1705.02695
  [cond-mat.mes-hall]} \BibitemShut {NoStop}%
\bibitem [{\citenamefont {{Huang}}\ \emph {et~al.}(2017)\citenamefont
  {{Huang}}, \citenamefont {{Zhou}},\ and\ \citenamefont
  {{Shen}}}]{Huang2017mwsm}%
  \BibitemOpen
  \bibfield  {author} {\bibinfo {author} {\bibfnamefont {Z.-M.}\ \bibnamefont
  {{Huang}}}, \bibinfo {author} {\bibfnamefont {J.}~\bibnamefont {{Zhou}}}, \
  and\ \bibinfo {author} {\bibfnamefont {S.-Q.}\ \bibnamefont {{Shen}}},\
  }\bibfield  {title} {\enquote {\bibinfo {title} {{Topological responses from
  chiral anomaly in multi-Weyl semimetals}},}\ }\href@noop {} {\bibfield
  {journal} {\bibinfo  {journal} {ArXiv e-prints}\ } (\bibinfo {year}
  {2017})},\ \Eprint {http://arxiv.org/abs/1705.04576} {arXiv:1705.04576
  [cond-mat.mes-hall]} \BibitemShut {NoStop}%
\bibitem [{\citenamefont {Ruderman}\ and\ \citenamefont
  {Kittel}(1954)}]{Ruderman1959rkky}%
  \BibitemOpen
  \bibfield  {author} {\bibinfo {author} {\bibfnamefont {M.~A.}\ \bibnamefont
  {Ruderman}}\ and\ \bibinfo {author} {\bibfnamefont {C.}~\bibnamefont
  {Kittel}},\ }\bibfield  {title} {\enquote {\bibinfo {title} {Indirect
  exchange coupling of nuclear magnetic moments by conduction electrons},}\
  }\href {\doibase 10.1103/PhysRev.96.99} {\bibfield  {journal} {\bibinfo
  {journal} {Phys. Rev.}\ }\textbf {\bibinfo {volume} {96}},\ \bibinfo {pages}
  {99--102} (\bibinfo {year} {1954})}\BibitemShut {NoStop}%
\bibitem [{\citenamefont {Kasuya}(1956)}]{kasuya1956rkky}%
  \BibitemOpen
  \bibfield  {author} {\bibinfo {author} {\bibfnamefont {Tadao}\ \bibnamefont
  {Kasuya}},\ }\bibfield  {title} {\enquote {\bibinfo {title} {A theory of
  metallic ferro-and antiferromagnetism on zener's model},}\ }\href@noop {}
  {\bibfield  {journal} {\bibinfo  {journal} {Progress of theoretical physics}\
  }\textbf {\bibinfo {volume} {16}},\ \bibinfo {pages} {45--57} (\bibinfo
  {year} {1956})}\BibitemShut {NoStop}%
\bibitem [{\citenamefont {Yosida}(1957)}]{Yosida1957rkky}%
  \BibitemOpen
  \bibfield  {author} {\bibinfo {author} {\bibfnamefont {Kei}\ \bibnamefont
  {Yosida}},\ }\bibfield  {title} {\enquote {\bibinfo {title} {Magnetic
  properties of cu-mn alloys},}\ }\href {\doibase 10.1103/PhysRev.106.893}
  {\bibfield  {journal} {\bibinfo  {journal} {Phys. Rev.}\ }\textbf {\bibinfo
  {volume} {106}},\ \bibinfo {pages} {893--898} (\bibinfo {year}
  {1957})}\BibitemShut {NoStop}%
\bibitem [{\citenamefont {Chang}\ \emph {et~al.}(2015)\citenamefont {Chang},
  \citenamefont {Zhou}, \citenamefont {Wang}, \citenamefont {Shan},\ and\
  \citenamefont {Xiao}}]{chang2015}%
  \BibitemOpen
  \bibfield  {author} {\bibinfo {author} {\bibfnamefont {Hao-Ran}\ \bibnamefont
  {Chang}}, \bibinfo {author} {\bibfnamefont {Jianhui}\ \bibnamefont {Zhou}},
  \bibinfo {author} {\bibfnamefont {Shi-Xiong}\ \bibnamefont {Wang}}, \bibinfo
  {author} {\bibfnamefont {Wen-Yu}\ \bibnamefont {Shan}}, \ and\ \bibinfo
  {author} {\bibfnamefont {Di}~\bibnamefont {Xiao}},\ }\bibfield  {title}
  {\enquote {\bibinfo {title} {{{RKKY}} interaction of magnetic impurities in
  {{Dirac}} and {{Weyl}} semimetals},}\ }\href {\doibase
  10.1103/PhysRevB.92.241103} {\bibfield  {journal} {\bibinfo  {journal} {Phys.
  Rev. B}\ }\textbf {\bibinfo {volume} {92}},\ \bibinfo {pages} {241103}
  (\bibinfo {year} {2015})}\BibitemShut {NoStop}%
\bibitem [{\citenamefont {Hosseini}\ and\ \citenamefont
  {Askari}(2015)}]{hosseini2015}%
  \BibitemOpen
  \bibfield  {author} {\bibinfo {author} {\bibfnamefont {Mir~Vahid}\
  \bibnamefont {Hosseini}}\ and\ \bibinfo {author} {\bibfnamefont {Mehdi}\
  \bibnamefont {Askari}},\ }\bibfield  {title} {\enquote {\bibinfo {title}
  {Ruderman-{{Kittel}}-{{Kasuya}}-{{Yosida}} interaction in {{Weyl}}
  semimetals},}\ }\href {\doibase 10.1103/PhysRevB.92.224435} {\bibfield
  {journal} {\bibinfo  {journal} {Phys. Rev. B}\ }\textbf {\bibinfo {volume}
  {92}},\ \bibinfo {pages} {224435} (\bibinfo {year} {2015})}\BibitemShut
  {NoStop}%
\bibitem [{\citenamefont {Dzyaloshinsky}(1958)}]{dzyaloshinsky1958dm}%
  \BibitemOpen
  \bibfield  {author} {\bibinfo {author} {\bibfnamefont {I}~\bibnamefont
  {Dzyaloshinsky}},\ }\bibfield  {title} {\enquote {\bibinfo {title} {A
  thermodynamic theory of ¡°weak¡± ferromagnetism of antiferromagnetics},}\
  }\href@noop {} {\bibfield  {journal} {\bibinfo  {journal} {Journal of Physics
  and Chemistry of Solids}\ }\textbf {\bibinfo {volume} {4}},\ \bibinfo {pages}
  {241--255} (\bibinfo {year} {1958})}\BibitemShut {NoStop}%
\bibitem [{\citenamefont {Moriya}(1960)}]{Moriya1960DM}%
  \BibitemOpen
  \bibfield  {author} {\bibinfo {author} {\bibfnamefont {T\^oru}\ \bibnamefont
  {Moriya}},\ }\bibfield  {title} {\enquote {\bibinfo {title} {Anisotropic
  superexchange interaction and weak ferromagnetism},}\ }\href {\doibase
  10.1103/PhysRev.120.91} {\bibfield  {journal} {\bibinfo  {journal} {Phys.
  Rev.}\ }\textbf {\bibinfo {volume} {120}},\ \bibinfo {pages} {91--98}
  (\bibinfo {year} {1960})}\BibitemShut {NoStop}%
\bibitem [{\citenamefont {Brey}\ \emph {et~al.}(2007)\citenamefont {Brey},
  \citenamefont {Fertig},\ and\ \citenamefont {Das~Sarma}}]{brey2007}%
  \BibitemOpen
  \bibfield  {author} {\bibinfo {author} {\bibfnamefont {L.}~\bibnamefont
  {Brey}}, \bibinfo {author} {\bibfnamefont {H.~A.}\ \bibnamefont {Fertig}}, \
  and\ \bibinfo {author} {\bibfnamefont {S.}~\bibnamefont {Das~Sarma}},\
  }\bibfield  {title} {\enquote {\bibinfo {title} {Diluted {{Graphene
  Antiferromagnet}}},}\ }\href {\doibase 10.1103/PhysRevLett.99.116802}
  {\bibfield  {journal} {\bibinfo  {journal} {Phys. Rev. Lett.}\ }\textbf
  {\bibinfo {volume} {99}},\ \bibinfo {pages} {116802} (\bibinfo {year}
  {2007})}\BibitemShut {NoStop}%
\bibitem [{\citenamefont {Saremi}(2007)}]{saremi2007}%
  \BibitemOpen
  \bibfield  {author} {\bibinfo {author} {\bibfnamefont {Saeed}\ \bibnamefont
  {Saremi}},\ }\bibfield  {title} {\enquote {\bibinfo {title} {{{RKKY}} in
  half-filled bipartite lattices: {{Graphene}} as an example},}\ }\href
  {\doibase 10.1103/PhysRevB.76.184430} {\bibfield  {journal} {\bibinfo
  {journal} {Phys. Rev. B}\ }\textbf {\bibinfo {volume} {76}},\ \bibinfo
  {pages} {184430} (\bibinfo {year} {2007})}\BibitemShut {NoStop}%
\bibitem [{\citenamefont {Gupta}(2017)}]{gupta2017}%
  \BibitemOpen
  \bibfield  {author} {\bibinfo {author} {\bibfnamefont {Amit}\ \bibnamefont
  {Gupta}},\ }\bibfield  {title} {\enquote {\bibinfo {title} {Floquet dynamics
  in multi-{{Weyl}} semimetals},}\ }\href {http://arxiv.org/abs/1703.07271}
  {\bibfield  {journal} {\bibinfo  {journal} {arXiv:1703.07271 [cond-mat]}\ }
  (\bibinfo {year} {2017})},\ \Eprint {http://arxiv.org/abs/1703.07271}
  {arXiv:1703.07271} \BibitemShut {NoStop}%
\bibitem [{\citenamefont {{Ezawa}}(2017)}]{Ezawa2017fmwsm}%
  \BibitemOpen
  \bibfield  {author} {\bibinfo {author} {\bibfnamefont {M.}~\bibnamefont
  {{Ezawa}}},\ }\bibfield  {title} {\enquote {\bibinfo {title} {{Photoinduced
  topological phase transition from a crossing-line nodal semimetal to a
  multiple-Weyl semimetal}},}\ }\href@noop {} {\bibfield  {journal} {\bibinfo
  {journal} {ArXiv e-prints}\ } (\bibinfo {year} {2017})},\ \Eprint
  {http://arxiv.org/abs/1705.02140} {arXiv:1705.02140 [cond-mat.mes-hall]}
  \BibitemShut {NoStop}%
\bibitem [{\citenamefont {{Yan}}\ and\ \citenamefont
  {{Wang}}(2017)}]{Yan2017fmwsm}%
  \BibitemOpen
  \bibfield  {author} {\bibinfo {author} {\bibfnamefont {Z.}~\bibnamefont
  {{Yan}}}\ and\ \bibinfo {author} {\bibfnamefont {Z.}~\bibnamefont {{Wang}}},\
  }\bibfield  {title} {\enquote {\bibinfo {title} {{Floquet multi-Weyl points
  in driven crossing-nodal-line semimetals}},}\ }\href@noop {} {\bibfield
  {journal} {\bibinfo  {journal} {ArXiv e-prints}\ } (\bibinfo {year}
  {2017})},\ \Eprint {http://arxiv.org/abs/1705.03056} {arXiv:1705.03056
  [cond-mat.str-el]} \BibitemShut {NoStop}%
\end{thebibliography}%

\end{document}